\documentclass{cimento}

\usepackage{graphicx}  
\usepackage{wasysym}
\title{Test beam results with prototypes for the new Cylindrical GEM Inner Tracker of the BESIII experiment}
\shorttitle{Test-beam results for the new Cylindrical GEM Inner Tracker of BESIII}

\author{L.~Lavezzi\from{ins:a}\from{ins:f}\thanks{Corresponding author ({\it e-mail:} lia.lavezzi@to.infn.it)}\ETC,
M. Alexeev\from{ins:f},
A. Amoroso\from{ins:f}\from{ins:l},
R. Baldini Ferroli\from{ins:a}\from{ins:c},
M. Bertani\from{ins:c},
D. Bettoni\from{ins:b},
F. Bianchi\from{ins:f}\from{ins:l},
A. Calcaterra\from{ins:c},
N. Canale\from{ins:b},
M. Capodiferro\from{ins:c}\from{ins:e},
V. Carassiti\from{ins:b},
S. Cerioni\from{ins:c},
JY. Chai\from{ins:a}\from{ins:f}\from{ins:h},
S. Chiozzi\from{ins:b},
G. Cibinetto\from{ins:b},
F. Cossio\from{ins:f}\from{ins:h},
A. Cotta Ramusino\from{ins:b}, 
F. De Mori\from{ins:f}\from{ins:l},
M. Destefanis\from{ins:f}\from{ins:l},
J. Dong\from{ins:c},
F. Evangelisti\from{ins:b},
R. Farinelli\from{ins:b}\from{ins:i},
L. Fava\from{ins:f},
G. Felici\from{ins:c},
E. Fioravanti\from{ins:b},
I. Garzia\from{ins:b}\from{ins:i},
M. Gatta\from{ins:c},
M. Greco\from{ins:f}\from{ins:l},
CY. Leng\from{ins:a}\from{ins:f}\from{ins:h},
H. Li\from{ins:a}\from{ins:f},
M. Maggiora\from{ins:f}\from{ins:l},
R. Malaguti\from{ins:b},
S. Marcello\from{ins:f}\from{ins:l},
M. Melchiorri\from{ins:b},
G. Mezzadri\from{ins:b}\from{ins:i},
M. Mignone\from{ins:f},
G. Morello\from{ins:c},
S. Pacetti\from{ins:d}\from{ins:k},
P. Patteri\from{ins:c},
J. Pellegrino\from{ins:f}\from{ins:l},
A. Pelosi\from{ins:c}\from{ins:e},
A. Rivetti\from{ins:f},
M. D. Rolo\from{ins:f},
M. Savri\'e\from{ins:b}\from{ins:i},
M. Scodeggio\from{ins:b}\from{ins:i},
E. Soldani\from{ins:c},
S. Sosio\from{ins:f}\from{ins:l},
S. Spataro\from{ins:f}\from{ins:l},
E. Tskhadadze\from{ins:c}\from{ins:g},
S. Verma\from{ins:i},
R. Wheadon\from{ins:f}
\atque 
L. Yan\from{ins:f} \\
\, \\
for the CGEM-IT group}

\instlist{\inst{ins:a} Institute of High Energy Physics, Chinese Academy of Sciences - Beijing, China
\inst{ins:b} INFN, Sezione di Ferrara - Ferrara, Italy
\inst{ins:c} INFN, Laboratori Nazionali di Frascati - Frascati (Roma), Italy
\inst{ins:d} INFN, Sezione di Perugia - Perugia, Italy
\inst{ins:e} INFN, Sezione di Roma, c/o Universit\`a La Sapienza - Roma, Italy
\inst{ins:f} INFN, Sezione di Torino - Torino, Italy
\inst{ins:g} Joint Institute for Nuclear Research (JINR) - Dubna, Russia
\inst{ins:h} Politecnico di Torino, Dipartimento di Elettronica e Telecomunicazioni - Torino, Italy
\inst{ins:i} Universit\`a di Ferrara, Dipartimento di Fisica - Ferrara, Italy
\inst{ins:k} Universit\`a di Perugia, Dipartimento di Fisica e Geologia - Perugia, Italy
\inst{ins:l} Universit\`a di Torino, Dipartimento di Fisica - Torino, Italy}

\begin{document}

\maketitle

\begin{abstract}
A cylindrical GEM tracker is under construction in order to replace and improve the inner tracking system of the BESIII experiment. Tests with planar chamber prototypes were carried out on the H4 beam line of SPS (CERN) with muons of $150$ GeV/{\it c} momentum, to evaluate the efficiency and resolution under different working conditions. The obtained efficiency was in the $96 - 98\%$ range. Two complementary algorithms for the position determination were developed: the charge centroid and the $\mu-$TPC methods. With the former, resolutions $<$$100$ $\mu$m and $<$$200$ $\mu$m were achieved without and with magnetic field, respectively. The $\mu-$TPC improved these results. By the end of 2016, the first cylindrical prototype was tested on the same beam line. It showed optimal stability under different settings. The comparison of its performance with respect to the planar chambers is ongoing. Here, the results of the planar prototype tests will be addressed.
\end{abstract}
\section{Introduction}
The BESIII experiment is taking data since 2009 at the BEPCII e$^+$e$^-$ collider in Beijing, PRC. A description of the machine and a summary of the most recent physics results are in \cite{besiii,bepcii,ifae2017-giulio}. \\
The current tracking system, a drift chamber, underwent a relevant radiation dose during the years, due to the growing luminosity up to the design value of $1 \cdot 10^{33}$ cm$^{-2}$ s$^{-1}$ in 2016 and an aging issue arose \cite{aging}. To cope with it, since the data taking is foreseen to last until 2022, with the option to continue up to 2027, the decision to replace the inner tracker with three layers of cyindrical triple-GEM was taken (CGEM-IT).
\section{The GEM detector}
The CGEM-IT consists of three layers of cylindrical triple-GEM. The GEM ({\bf G}as {\bf E}lectron {\bf M}ultiplier) was invented in 1997 by F. Sauli \cite{sauli} to achieve a high multiplication gain in gas with a low discharge rate and minimizing the problem of positive ion backflow. The GEM foil is a metal coated polymer ($50+3$ $\mu$m kapton/copper), pierced with thousands of holes ($\diameter$ $50$ $\mu$m). The electric field $\sim$tens of kV/cm to produce the electronic avalanche of the ionization electrons is not set by wires, as in standard gas detectors, but through the tiny holes, by applying a voltage of some hundreds of Volts between the copper surfaces. A gain around $10^4$ is achievable with moderate voltages and evan higher values can be obtained by placing three GEM foils instead of just one between anode and cathode and operating them at lower voltages, thus decreasing the discharge probability \cite{triple}. \\
Together with the restoration of the original inner tracker efficiency, the installation of the CGEM-IT will provide an improvement in the charged track position determination. The two-views readout of the anodic plane, with strips both parallel to the beam direction and tilted of a large stereo angle ($> 30^\circ$), will permit the reconstruction of the tridimentional position of the particle passage on each of the three layers. In particular, this will improve the $z$ coordinate resolution, granting the same r-$\phi$ position and momentum resolutions of the current tracker. The reconstruction of short living particles will benefit from this. \\
For the detector development, the CGEM-IT exploits the experience of the KLOE-2 experiment (Frascati) which owns the first cylindrical triple-GEM, with important innovations. The Vertical Inserting Machine \cite{kloe2} which is used to assemble the cylindrical electrodes in a complete layer, was invented by KLOE-2 and modified for BESIII. \\
Various improvements have been added with the aim to obtain the best performances and satisfy the BESIII requirements in terms of material budget. Honeycomb was replaced by Rohacell 31 as support for anode and cathode for its lightness and robustness. The permaglass rings which hold the layers are placed only outside the active area. \\
The BESIII CGEM-IT will be the first cylindrical GEM with analog readout inside a magnetic field, with the simultaneous measurement of the deposited charge and the time of arrival of the signal. A dedicated ASIC has been designed and is under test (TIGER \cite{roloINSTR17}). Moreover, the jagged strip layout of the anode will lower the inter-strip capacitance \cite{isaTIPP14}. Finally, with a total X$_0 < 1.5\%$ and the capability to sustain the high particle rate ($\sim$$10^4$ Hz/cm$^2$), the CGEM-IT is not only a technological upgrade for BESIII, but will also improve the quality of the results under the scientific point of view.
\section{The position reconstruction: algorithms and test beam results}
\begin{figure}[h!tbp]
\centering 
\includegraphics[width=.8\textwidth]{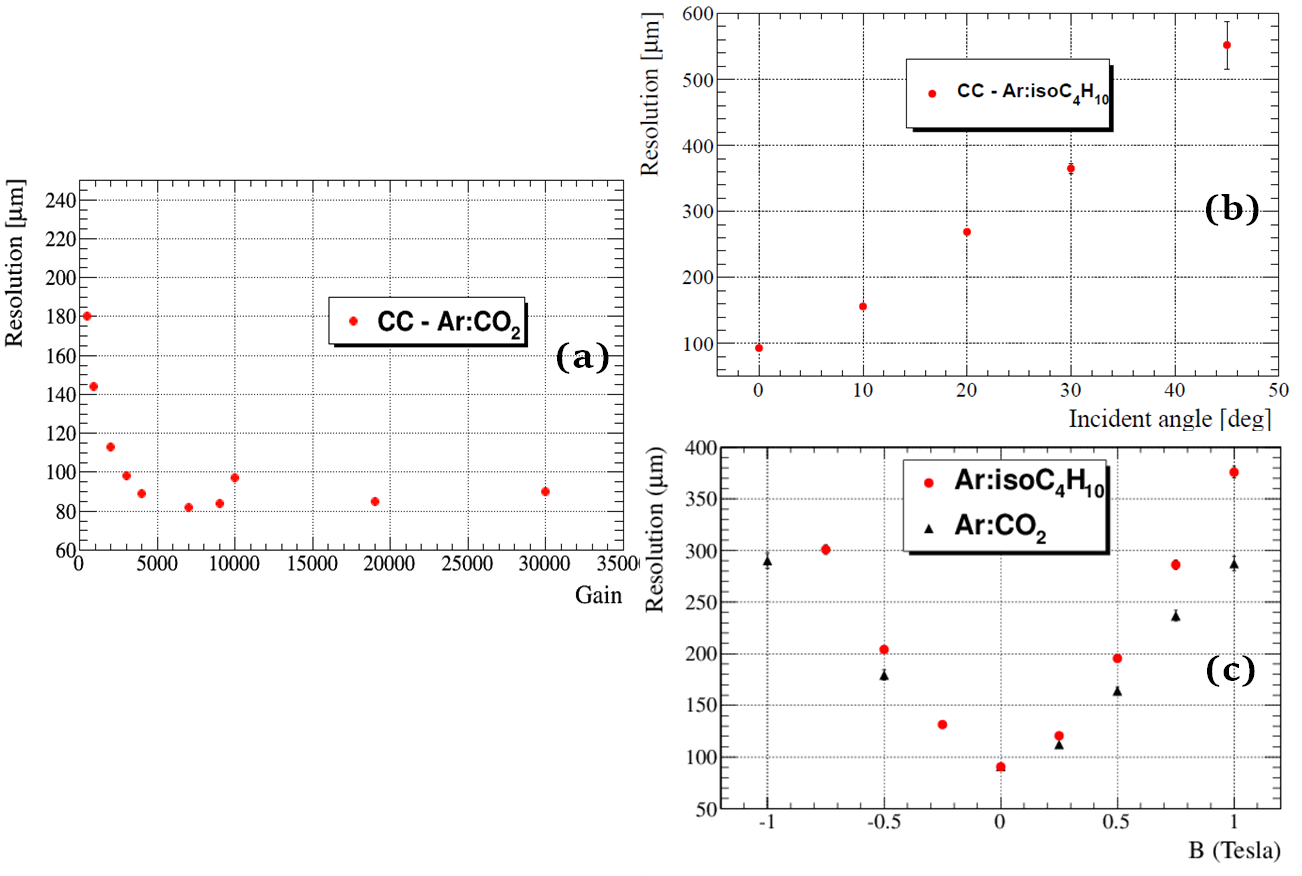} 
\caption{CC resolution: (a) orthogonal tracks without magnetic field, (b) inclined tracks without magnetic field and (c) orthogonal tracks inside a magnetic field.} \label{fig:res_cc}
\end{figure}

\begin{figure}[h!tbp]
\centering 
\includegraphics[width=.45\textwidth]{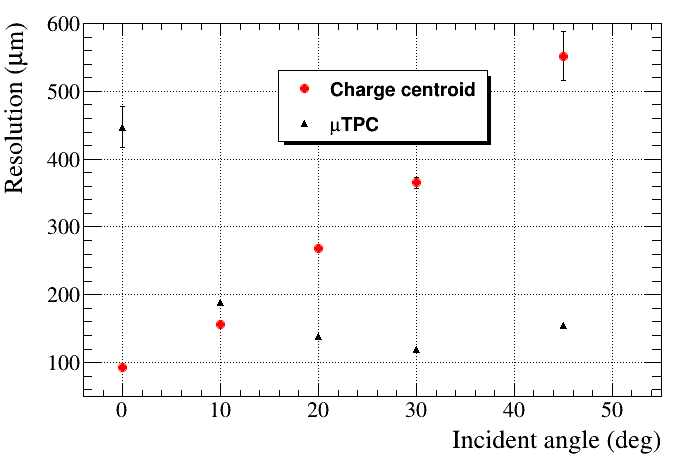} 
\qquad
\includegraphics[width=.45\textwidth]{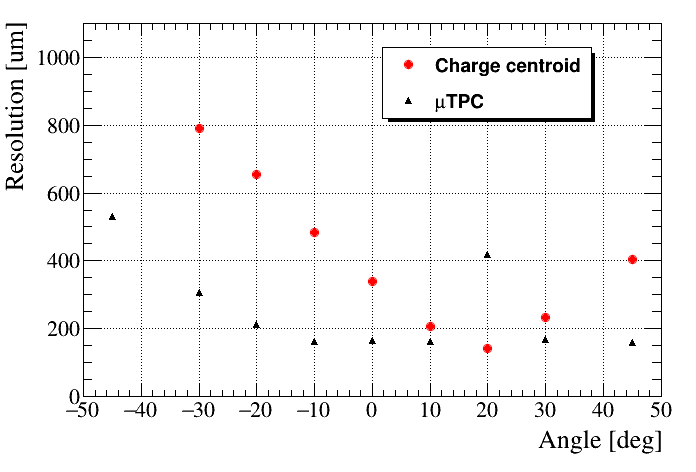}
\caption{Comparison between CC (red circles) and $\mu$-TPC (black tiangles) for angle $\neq 0$ and B $= 0$ ({\it left}) or B $\neq 0$ ({\it right}). The graphs were obtained in Ar/CO$_2$ (Lorentz angle is $20^\circ$) \cite{ricIEEE}.} \label{fig:res_cc_mutpc}
\end{figure}

Three scenarios are possible (from the simplest to the most complicate):
\begin{itemize}
\item[1.] magnetic field {\it off} and orthogonal incident tracks;
\item[2.] magnetic field {\it on} ({\it off}) and orthogonal (inclined) incident tracks;
\item[3.] magnetic field {\it on} and inclined tracks simultaneously.
\end{itemize}
Two position reconstruction algorithms have been developed: the charge centroid (CC) and the micro-TPC ($\mu-$TPC) methods. The former is {\it traditional} and suitable in case $1$, where the charge distribution on the anode is symmetric and gaussian. The latter has been developed for ATLAS MicroMegas \cite{mu-TPC} and takes advantage from the larger charge distribution on the strips in the case of inclined tracks and/or Lorentz force effect. Depending on the magnetic field intensity and the incident angle entity, the two methods show complementary performances. \\
Three plots of resolution are shown in fig.\ref{fig:res_cc} to highlight the difference between case $1$ and $2$. They have been obtained with $10 \times 10$ cm$^2$ planar chambers in tests performed at the H4 beam line of SPS (CERN) changing the working settings (more details in \cite{ricIEEE,liaINSTR17}). 
The most challenging situation is when the Lorentz force (B $\neq 0$) and the inclination angle act together. {\it Focusing} and {\it de-focusing} effects show up, depending on the concordance or discordance between Lorentz and track inclination angles. The application of the CC and the $\mu-$TPC modes have specular behaviours (see fig.\ref{fig:res_cc_mutpc}). A proper combination of the two calculations will provide the requested resolution of $130$ $\mu$m on the full BESIII angular range and inside the experimental magnetic field of $1$T.

\acknowledgments
The research leading to these results has been performed within the BESIIICGEM Project, funded by the European Commission in the call H2020-MSCA-RISE-2014.


\vspace{4cm}

\begin{thebibliography}{99}

\bibitem{besiii}
\BY{M. Ablikim et al.} \IN{Nucl. Instr. Meth. A}{614/3}{2010}{345-399}

\bibitem{bepcii}
\BY{C. Zhang, G.X. Pei} \IN{Proceedings of 2005 Particle Accelerator Conference}{}{2005}{131-135}

\bibitem{ifae2017-giulio}
\BY{G. Mezzadri et al.} these proceedings

\bibitem{aging}
\BY{MY. Dong et al.} \IN{Chinese Physics C}{40}{2016}{016001}

\bibitem{sauli}
\BY{F. Sauli} \IN{Nucl. Instr. and Meth. A}{386}{1997}{53l-534}

\bibitem{triple}
\BY{S. Bachmann et al} \IN{Nucl. Instr. and Meth. A}{479}{2002}{294}

\bibitem{kloe2}
\BY{S. Balla et al.} \IN{JINST}{9}{2014}{C01014}

\bibitem{roloINSTR17}
\BY{M. D. Da Rocha Rolo et al.} arXiv:1706.02267 [physics.ins-det]
  	
\bibitem{isaTIPP14}
\BY{D. Bettoni et al.} \IN{PoS(TIPP2014)}{292}{2014}{}

\bibitem{mu-TPC}
\BY{M. Iodice} \IN{JINST}{9}{2014}{C01017}

\bibitem{ricIEEE}
\BY{M. Alexeev et al.} to be published on \emph{IEEE 2016 Conference RECORD}

\bibitem{liaINSTR17}
\BY{L. Lavezzi et al.} arXiv:1706.02428 [physics.ins-det]

\end{thebibliography}
\end{document}